\documentclass[journal]{IEEEtran}
\usepackage{epsfig,color,amsmath,cite}

\IEEEoverridecommandlockouts
\usepackage{wrapfig}
\usepackage{bm}
\usepackage{epstopdf}
\usepackage{amssymb}
\usepackage{url}
\usepackage{enumitem}
\usepackage{multirow}
\usepackage{booktabs}
\usepackage{makecell}
\usepackage{algorithm,algorithmic}	
\setlist[itemize]{leftmargin=*}

\DeclareMathOperator*{\minimize}{minimize}

\DeclareMathOperator{\subjectto}{subject\ to}

\makeatother

\newcommand{\st}{{\rm s.t.}}

\newcommand{\m}{\boldsymbol}
\allowdisplaybreaks[4]
\usepackage[colorlinks = true,
linkcolor = blue,
urlcolor  = blue,
citecolor = blue,
anchorcolor = blue]{hyperref}

\newcommand{\mc}[1]{\mathcal{#1}}

\usepackage[framemethod=TikZ]{mdframed}
\mdfdefinestyle{MyFrame}{%
	linecolor=black,
	outerlinewidth=1pt,
	roundcorner=4pt,
	innerrightmargin=5pt,
	innerleftmargin=5pt,}

\begin{document}
\title{\vspace{-0.53cm}Blockchain-Assisted Crowdsourced Energy Systems}
\author{Shen Wang*, Ahmad F. Taha*, and Jianhui Wang**
\thanks{*Department of Electrical and Computer Engineering,	The University of Texas at San Antonio, Texas 78249. **Department of Electrical Engineering, Southern Methodist University,	Dallas, Texas 75205. Emails: \texttt{mvy292@my.utsa.edu, ahmad.taha@utsa.edu, jianhui@smu.edu}. This work will be presented at the IEEE Power \& Energy Society General Meeting 2018 in Portland, Oregon, August 5--10, 2018. }\vspace{-1cm}}


\maketitle

\begin{abstract}

Crowdsourcing relies on people's contributions to meet product- or system-level objectives. Crowdsourcing-based methods have been implemented in various cyber-physical systems and realtime markets.
This paper explores a framework for Crowdsourced Energy Systems (CES), where small-scale energy generation or energy trading is crowdsourced from distributed energy resources, electric vehicles, and shapable loads. The merits/pillars of energy crowdsourcing are discussed. Then, an operational model for CESs in distribution networks with different types of crowdsourcees is proposed. The model yields a market equilibrium depicting traditional and distributed generator and load setpoints.  Given these setpoints, crowdsourcing incentives are designed to steer crowdsourcees to the equilibrium. As the number of crowdsourcees and energy trading transactions scales up, a secure energy trading platform is required. To that end, the presented framework is integrated with a lightweight Blockchain implementation and smart contracts. Numerical tests are provided to showcase the overall implementation.


\end{abstract}
%

\vspace{-0.43cm}
\section{Introduction}~\label{sec:intro}
\vspace{-0.43cm}

C{rowdsourcing}, democratized marketplaces, and collaborative production are all attributes of the sharing economy. People's drive for a self-sustainment has driven the world to this disrupting economy. By 2025, spending of the US sharing economy is projected to exceed \$335 billion~\cite{pwc}.
The democratization and evolution of the world markets in general, and potentially future energy systems, necessitates novel crowdsourcing-based methodologies to deal with this evolution.
This paper puts forth the high-level implementation of crowdsourcing design mechanisms for collaborative production and consumption in energy markets that are reliant on small-scale distributed generation with the assistance of cryptocurrencies and Blockchain-assisted energy systems operation---leading to
\emph{{Crowdsourced Energy Systems}} (CES).



\subsection{Crowdsourced Energy Systems: A Simple Example}

Crowdsourcing is a major drive for various industries, and has been utilized in various disciplines such as medicine, cyber physical systems, and engineering system design. The central theme in crowdsourcing is the utilization of the crowd's power to achieve either product-level or system-level objectives.

To understand how crowdsourcing can be applied in energy systems, we provide an analogy from the Internet's most popular crowdsourcing markets, the Amazon Mechanical Turk (MTurk). MTurk enables people to post jobs with monetary rewards and expiry dates. The tasks in CESs can be plugging in an electric vehicle, charging/discharging a battery (at a certain rate), and supplying the power network with renewable energy via solar panels---with the objective of satisfying a demand shortage. These tasks can be automated via smart inverters, plugs, and meters, while interfacing with power utilities.

The human component is manifested through the interaction between the \textit{crowdsourcer} (CES operator or utility) and the \textit{crowdsourcee} (energy producer). Some of these tasks have urgent expiry deadlines, while others are predetermined ahead of time as in day-ahead markets.
Energy-aware people can then accept or decline the proposed transactions. The utility's objective is to guarantee that any computed, realtime market equilibrium is robust enough to a large percentage of crowdsourcees declining the transactions---while guaranteeing the grid's transient stability and operational constraints as well as satisfying environmental regulations.
In the future, with the transformation in the role of utilities, these transactions can be encrypted through smart contracts via the Blockchain technology with or without cryptocurrencies~\cite{BCSurvey2017}.
\subsection{Related Literature}

Recent research studies have investigated integrating the operation of distributed energy resources (DERs) in distribution networks and deregulated power systems. The focus of the majority of these studies is on unit commitment and economic dispatch problems as well as scheduling of DERs, given load and renewables uncertainty. The work in~\cite{wang2008security} investigates security-constrained unit commitment with volatile wind power. The authors in~\cite{guggilam2016scalable} present computationally tractable optimization routines to manage the operation distribution networks with high PV penetration. 


Another branch of related work studies the design of demand response signals and incentives to drive DER owners to contribute to energy production. The authors in~\cite{hajiesmaili2017crowd} investigate the problem of utilizing heterogeneous, crowdsourced energy storage systems in microgrids to perform demand response. In summary, there are two major approaches to demand response: (a) Reducing the total demand by either using their local DERs or shifting demand as the grid requires; (b) Designing robust generator setpoints to reduce the total generation while considering worst-case scenarios~\cite{DR1}. In this paper, we focus on the first kind of demand response schedules, in addition to the design of monetary crowdsourcing incentives to steer DERs to a certain market equilibrium.

Section~\ref{sec:CS} discusses the merits of energy crowdsourcing. Section~\ref{sec:C1} presents an integrated optimal power flow optimization model for CESs. The solution computed from this optimization model is referred to as the market equilibrium that includes generator setpoints. Given the setpoints, Section~\ref{sec:C3} proposes simple crowdsourcing incentives to steer crowdsourcees to the computed equilibrium. Section~\ref{sec:BC} explores a Blockchain application of this framework. Section~\ref{sec:tests} concludes with simple numerical tests.

\section{Energy Crowdsourcing Merits and Pillars}~\label{sec:CS}

We outline three merits of studying crowdsourcing in smart grids.  First, power generation shifts from being centralized and fuel-dependent to distributed, small-scale, and renewables-dependent. Utilities can crowdsource the time-critical energy generation from consumers and loads with DERs, when utilities cannot meet the demand or the minimum percentage of renewable penetration.
the latter consequently empowers people to be more proactive towards energy systems issues, which can in turn raise awareness and increase relevance of energy systems challenges.

Second, in its essence, crowdsourcing mechanisms explore the design of incentives to steer people or users to perform certain duties, while expanding the set of distributed generation contributions.
This can take the form of demand response in power systems through dynamics-aware, distributed pricing schemes that ensure the grid's stability while minimizing costs. Third, the growth of Blockchain and cryptocurrencies can speed up the elimination of the so-called \textit{middle-man} in energy systems, by facilitating secure peer-to-peer energy transactions.
In fact, utilities in Austria have been experimenting with energy trading via Blockchain, and solar panel owners in Brooklyn are energy trading via the same platform \cite{Cottrell_2017}. However, these trades cannot scale-up without a computational framework that ensures reliability of smart grids. 

The focus of this paper is on establishing a framework that guides the exponential increase of such crowdsourced transactions while not being restricted to cryptocurrencies.
\begin{figure}[t]
		\centering \includegraphics[scale=0.25]{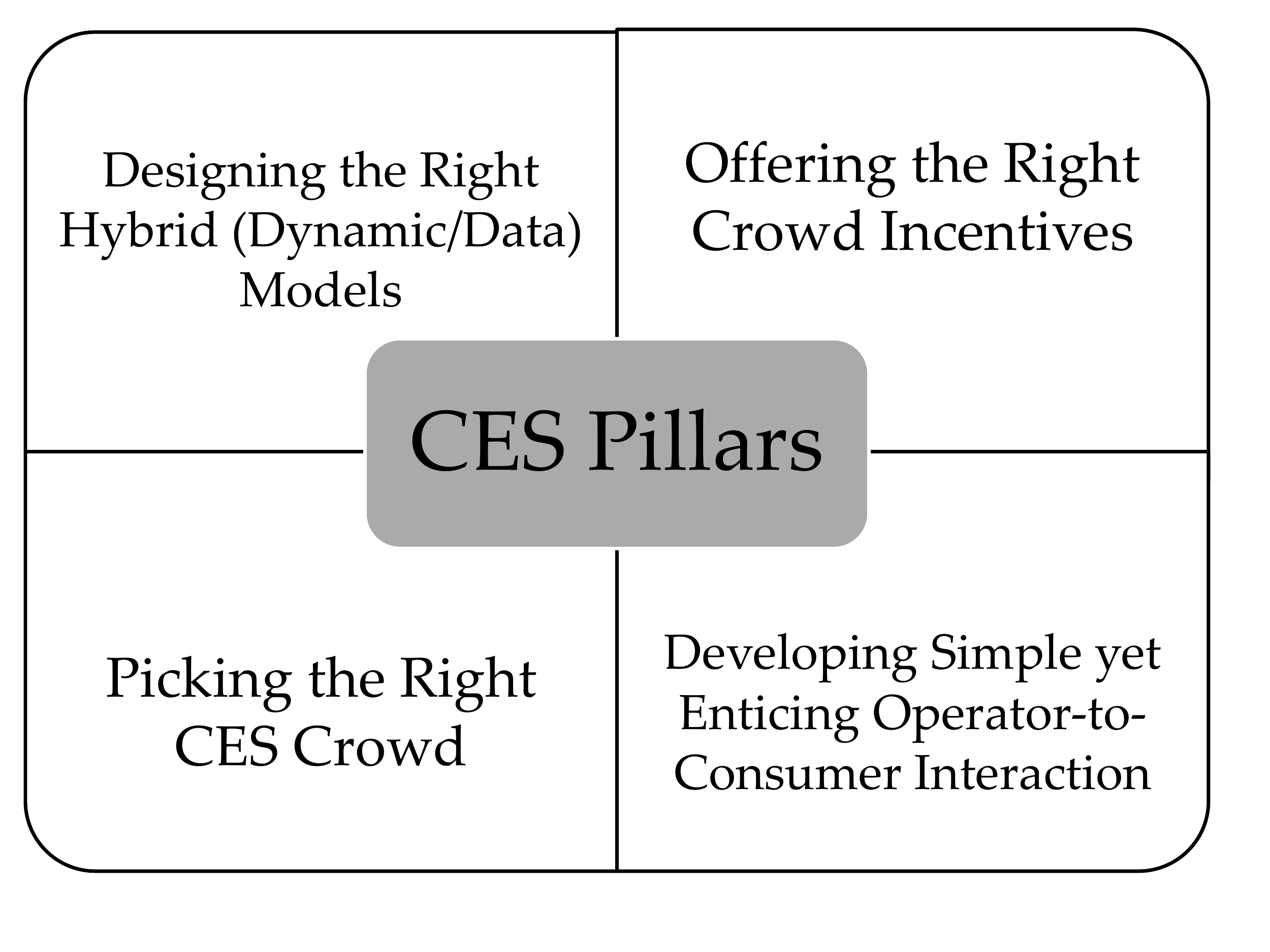}
	\vspace{-0.4cm}
	\caption{The major components of crowdsourcing-driven energy systems.}
	\label{pillars}
\vspace{-0.55cm}
\end{figure}
In~\cite{Howe2008}, the author presents the short history and ambitious future of crowdsourcing. The main pillars of successful crowdsourcing in energy systems are depicted in Fig.~\ref{pillars}, for CESs. For brevity, we do not discuss these pillars in depth here.

\section{Integrated Operational Model of CESs}~\label{sec:C1}

We model a distribution network by a graph $\mathcal{T}:= (\mathcal{N},\mathcal{E})$. The root of the tree $\mathcal{T}$ is a substation bus that is connected to the transmission network. A feeder connects the substation to this distribution area.
Define the partition $\mathcal{N} = \mathcal{G} \bigcup \mathcal{C} \bigcup \mathcal{L} $, where $\mathcal{G}$ collects the buses with a generator; $\mathcal{C} = \{1,\ldots, n_c\}$ collects the buses containing $n_c$ agents with DERs who signed up for crowdsourcing schedules; $\mathcal{L} = \{1,\ldots, n_l\}$ collects $n_l$ load buses.
The \textit{crowdsourcer} here is the utility company or any other system operator. We distinguish between two types of DERs in $\mc{C}$, also referred to as \textit{crowdsourcees}. The first type commits in the day-ahead market (and perhaps monthly or yearly) to the crowdsourcing tasks requested by the system operator.
Type 2 crowdsourcees provide near realtime adjustments or decisions based on realtime notifications and decisions from the utility.
The distinction between these two types of users is needed as it resembles projected market setups. We define these two types as $\mc{C}_{T1}$ and $\mc{C}_{T2}$.

\subsection{CES Optimal Power Flow with Distributed Generation}


Let $i$ denote the bus index of the distribution system and $t$ denote the time-period. In this paper, we consider bulk, dispatchable generation from traditional synchronous generators, renewable energy generation from solar panels, fully controllable stationary batteries, uncontrollable loads, and {shapable} loads. Due to the space limitation, we only linguistically describe the operational setup instead of including the full optimization models of batteries, uncontrollable loads and shapable loads. The detailed models are common in the literature~\cite{taylor2015convex,Munsing2017}.

\subsubsection{Solar Energy Generation}

Solar panels generate real power $p_{i,t}^r$ at bus $i \in \mc{R} \subset \mc{C}$ ($\mc{R}$ is the set collecting the buses with solar panels) in time-period $t$.
Note that Type 1 crowdsourcees do not control whether $p_{i,t}^{r}$ is fed into the grid or not, whereas Type 2 crowdsourcees dictate whether to use $p_{i,t}^r$ locally or sell it to the CES operator or other users.
\subsubsection{Stationary Batteries}
Batteries are modeled as dispatchable loads that can withdraw or inject power denoted as $p_{i,t}^b$ for $i \in \mc{B} \subset \mc{C}$ ($\mc{B}$ is the set collecting the buses with batteries).

\subsubsection{Uncontrollable and Shapable Loads} Uncontrollable loads (lights, plug loads, street lights, etc \ldots) are considered to be given and are denoted by $s_{i,t}^u$ for all $i\in \mc{L}$ (loads can include reactive power). We also consider shapable loads, defined by $s_{i,t}^s$ for $i\in \mc{L}_s \subset \mc{L}$ ($\mc{L}_s$ is the set collecting the buses with shapable loads), such as electric vehicles with flexible power profile but fixed energy demand in 24 hours. These loads
must be charged during a certain period of time. 

\subsubsection{Dispatchable Generators} Dispatchable generators are considered in this paper with a quadratic cost functions
$C_{i,t}(p_{i,t}^g)$ in terms of the real power generated $p_{i,t}^g$ for $i \in \mc{G}$. 


\subsubsection{Branch Flow Model and Operational Constraints}
For the above types of generators and loads, we consider that they all have certain linear operational constraints denoted by $\mathcal{X}$ (see~\cite{taylor2015convex} for concrete battery and shapable loads models). For Type 2 crowdsourcees, these operational constraints or energy usage preferences are communicated ahead of time.
For all buses in the network, define $ \m {x_1}:=(\m{p^b},\m{p^r},\m{s^g},\m{s^u}, \m{s^s})$ as the vector collecting the apparent power output of DERs, loads, and generators.
To model power flow in distribution networks, we adopt a branch flow model first proposed in~\cite{baran1989optimal}. For simplicity, all of the branch flow variables are included in a single variable $ \m {x_2}:=(\m{v},\m{l},\m{P},\m{Q},\m{p},\m{q})$.  The branch flow model is denoted by $\mathrm{BranchFlowModel}({\m x_2})$ in terms of $\m x_2$.
\subsubsection{Overall Optimization Problem}
The overall CES Optimal Power Flow (CES-OPF) can be compactly written as:
\vspace{-0.25cm}
{\begin{subequations}~\label{equ:OPF}
	\begin{eqnarray}
	\hspace{-0.75cm}\textbf{CES-OPF:}\;\;
	\minimize_{\m x_{1_{t}}, \m x_{2_{t}}} && \hspace{-0.3cm}\sum_{t=1}^{T} C_t (\m {x_{1_{t}}},\m {x_{2_{t}}})\\
	\subjectto && \m x_{1_{t}} \in \mathcal{X} \\
	&& \mathrm{BranchFlowModel}({\m x_{2_{t}}}). \label{equ:BF}
	\end{eqnarray}
\end{subequations}}
The optimization~\eqref{equ:OPF} minimizes the cost functions of traditional generators and other user-defined cost functions on the battery operation or the shapable loads for all crowdsourcees, subject to the branch flow model, operational and thermal limit constraints. Using the second order cone program (SOCP) relaxation of the branch flow model~\cite{farivar2013branch}, CES-OPF can formulated as a convex program with SOCP constraints. 

The CES-OPF can be decomposed into small optimization subproblems by decoupling variables and constraints---the overall problem can be then solved through a decentralized ADMM algorithm; see~\cite{peng2014distributed}. Another approach is to simply solve CES-OPF in a centralized fashion. For medium- or small-scale distribution networks, it is plausible to solve CES-OPF in a centralized fashion after requesting the user's preferences---denoted by $\mathcal{X}$---ahead of time, as an alternative to implementing the decentralized ADMM algorithm.

\section{ Crowdsourcing Incentives}~\label{sec:C3}
As outlined in Section~\ref{sec:C1}, we solve CES-OPF and obtain the equilibrium (or setpoints) for the three major energy producers: utility-scale power plants, Type 1 and Type 2 crowdsourcees.
Due to the nature of contracts with traditional generators and Type 1 crowdsourcees, the CES operator controls all of these decision variables except $p_{i}^r, p_i^b, p_{i}^u$ for $i\in \left(\mc{B} \bigcup \mc{R} \bigcup \mc{L}_s\right) \bigcap \mc{C}_{T2}$ meaning that additional incentives are to be provided for Type 2 crowdsourcees to accept the solutions of~\eqref{equ:OPF} in realtime markets. Here, we outline the design of economic incentives to steer Type 2 crowdsourcees to the equilibrium solution.
The problem of designing realtime pricing and incentives is a recent one in the smart grid literature
\cite{JOKIC2009,langbort2010real,Namerikawa2015}. These methods fuse simple game- and control-theoretic frameworks to obtain optimal pricing and operation with regulation. Here, we use a different approach to solve for monetary incentives.

The distributed locational marginal price (DLMP) vector $\m \lambda_t$ is computed at time-period $t$ by finding the dual variables associated with the power balance equations in branch flow model~\eqref{equ:BF}. We consider that Type 2 crowdsourcees at bus $i$ receive $\lambda_{i,t} + \lambda_{i,t}^a$ where $\lambda_{i,t}^a$ is the additional monetary reward---an optimization variable that we solve for.
Note that the equilibrium of the first two energy producers is dispatchable through smart grid communication technologies.


The crowdsourcing incentives are defined at the local community level in a distribution system.
We also assume that (a) the operator has allocated a budget to spend on the realtime incentives at the feeder level with a budget range $[b_t^{\min} \;,\;  b_t^{\max}]$, and (b) there is a demand shortage or surplus $d_t$. Assuming that $u_i^{\mathrm{eq.}}$  (which includes $p_{i}^r, p_i^b,$ or  $p_{i}^u$ for each bus) is the equilibrium computed by \eqref{equ:OPF} for $i \in \mc{C}_{T2}$, the proposed CES Incentive Design (CES-ID) routine at time $t$ is formulated as:
\vspace{-0.36cm}
{\begin{subequations}~\label{equ:incentives}
	\begin{align}
\hspace{-0.55cm}\textbf{CES-ID:}\;\min_{\m \lambda^a,\m u^a} &\;\; \sum_{i =1}^{|\mc{C}_{T2}|} \eta_i \left(\lambda_{i,t}+\lambda_{i,t}^a\right) u_{i,t}^a + \zeta_i |u_{i,t}^a - u_{i,t}^{\mathrm{eq.}} |\\
\st& \;\;b_t^{\min} \leq \sum_{i =1}^{|\mc{C}_{T2}|} \lambda_{i,t}^a u_{i,t}^a  \leq b_t^{\max} \\
& \;\;\sum_{i =1}^{|\mc{C}_{T2}|} u_{i,t}^a  \geq d_t,\;\; u_{i,t} \in \mc{U}_{i,t}^a, \; \lambda_{i,t}^a > 0.
\end{align}
\end{subequations}}
In~\eqref{equ:incentives}, $\m \lambda^a$ and $\m u^a$ are the vector variables collecting the monetary incentives and the generation quantities for $i \in \mc{C}_{T2}$; $u_i^{\mathrm{eq.}}$ is the equilibrium solution from CES-OPF; $\zeta_i$ and $\eta_i$ are crucial weights that are obtained based on the historic preferences of crowdsourcee $i$ and their willingness to accept the offers; $\mc{U}_i^a$ depicts linear operational constraints of the network and Type 2 crowdsourcees. The objective of CES-ID is to steer the crowdsourcees to accept the market equilibrium computed in~\eqref{equ:OPF} through the budgeted incentives.

Problem~\eqref{equ:incentives} is non-convex due to the term $\lambda_i^a u_i^a$. This routine can be relaxed to a linear problem after substituting $y_i=\lambda_i^a u_i^a$ in the objective and constraints, and adding $y_i>0$ as a necessary constraint. This linear relaxation is an adaptation of the {reformulation linearization technique}~\cite{sherali1992global}. Given this linearization, \eqref{equ:incentives} is a tractable linear program that fits the need of realtime crowdsourcing and ancillary services of Type 2 crowdsourcees.

\vspace{-0.2cm}
\section{Blockchain and Smart Contracts for CESs}~\label{sec:BC}
\vspace{-0.3cm}

\begin{table}
	\fontsize{7.5}{7}\selectfont
	\caption{Various Implementations of Blockchain. PoW and PBFT stand for Proof of Work and Practical Byzantine Fault Tolerance.}
	\centering
	\makegapedcells
	\setcellgapes{0.85pt}
		\begin{tabular}{ c|c|c|c }
		& \textit{Bitcoin} & \textit{Ethereum} & \textit{Hyperledger Fabric} \\
		\hline
	{\textit{Year Released}}
		& 2009 & 2015 & 2017 \\
		\hline
		\textit{Cryptocurrency} & bitcoin & ether & none \\
		\hline
		\textit{Network}  & public & public/permissioned & permissioned \\ 		\hline
		\textit{Transactions} & anonymous & anonymous/private & public/confidential \\		\hline
		\textit{Consensus} & PoW & PoW & PBFT \\ 		\hline
		\textit{Smart Contracts}	& None & Solidity/Serpent & Chaincode \\ 		\hline
		\textit{Language} &	C++ & C++/Python &	Golang/Java \\
		\hline		\hline
	\end{tabular}
	\label{table:1}
\end{table}

\vspace{-0.53cm}
\subsection{Blockchain for Crowdsourced Energy Systems}
Blockchain is a distributed peer-to-peer, computational database that underlines and facilitates transactions through cryptocurrencies such as Bitcoin and Ether~\cite{swan2015blockchain}; Blockchain technology can also be used with traditional currencies through tokens. The main components in a Blockchain are interlinked, secure, and time-stamped \textit{blocks} that define transactions between users.
The main motivation behind Blockchain is the need to have a distributed, secure system that eliminates the need for the so-called \textit{middle man or a central authority} that organizes transactions.
Novel cryptographic techniques are central to Blockchain, ensuring that any trade of any commodity between two users is securely replicated throughout all decentralized databases. Through cryptocurrencies, Blockchain ensures that transactions cannot be forged.
The reader is referred to~\cite{swan2015blockchain} for an in depth study on Blockchain, Ethereum, and Bitcoin with applications in various industries.

Smart contracts---protocols developed to verify the performance of a contract---are a major component of Blockchain. Smart contracts and Blockchain provide an excellent platform to perform energy trading transactions. In particular, the authors in~\cite{BCSurvey2017,licata_2017} provide a high-level description to the main merits of using cryptocurrencies and Blockchain in energy systems. Managing the contracts through Blockchain is favored, as the system operator has many incentives to manage the transactions via a secure, third party. This, unlike other Blockchain applications, still requires a central authority---the utility company or the system operator managing the grid. Small-scale energy trading without a central authority can take place (see~\cite{Cottrell_2017}), yet the scaling of these transactions to include thousands of people and millions of daily energy transactions without the utility interfering is remote in today's markets. To that end, the proposed architecture in this paper still requires a central authority to manage the grid.

The authors in~\cite{Hahn2017} design a smart contract to enable energy producers to sell excess energy to the highest bidder through an auction which incentives honest bidding. Through the Ethereum Blockchain, this design is tested at Washington State University campus. The study in~\cite{Horta2016} investigates a paradigm for providing demand-side management services where households can manage their surplus/shortage of energy through self-enforcing smart contracts via Blockchain. The authors in~\cite{Munsing2017} develop a novel Blockchain-based architecture for peer-to-peer energy markets which includes a decentralized optimal power flow formulations, similar to~\eqref{equ:OPF}, that is then envisioned through smart contracts. Recently, a Blockchain-based transactive energy systems implementation that preserves privacy of users is proposed in~\cite{kvaternik2017privacy}.

\vspace{-0.34cm}
\subsection{Blockchain Implementations and Energy Inefficiency}
Table~\ref{table:1} summarizes the attributes of different implementations of Blockchain. A consensus protocol is used to ensure the unambiguous ordering of transactions and guarantee the integrity and consistency of the Blockchain across distributed nodes~\cite{baliga2017understanding}; various consensus protocols
have been developed for Blockchain implementations. An important component of Blockchain is the \textit{mining} process where miners validate new transactions and record them on the global Blockchain ledger. In Proof of Work (PoW) consensus protocol, used by Bitcoin and Ethereum, miners race to add new transactions to the ledger by competing to solve a complex, computationally intensive and energy-consuming cryptographic puzzle.
When scaled, this would result in an energy inefficient platform that defeats the purpose of sustainable energy systems. In fact, the annual estimated electricity consumption of Bitcoin is 46.86 Terawatt-hour---a staggering 0.21\% of world’s electricity consumption~\cite{energyconsumption}. For any Blockchain implementation to be successful in CESs, the energy consumption of the consensus technology for each energy trading transaction should only require little energy.
To that end, we propose using the IBM Hyperledger Fabric~\cite{worldstate} that uses Practical Byzantine Fault Tolerance (PBFT) as its consensus protocol. PBFT consumes orders of magnitude less energy in comparison with other consensus protocols \cite{vukolic2015quest}. In addition, Hyperledger Fabric requires minimal user interaction as a simple, user-driven interactive front end can be developed using this technology.




\begin{figure}[h]
		\vspace{-0.2cm}
 \includegraphics[scale=0.33]{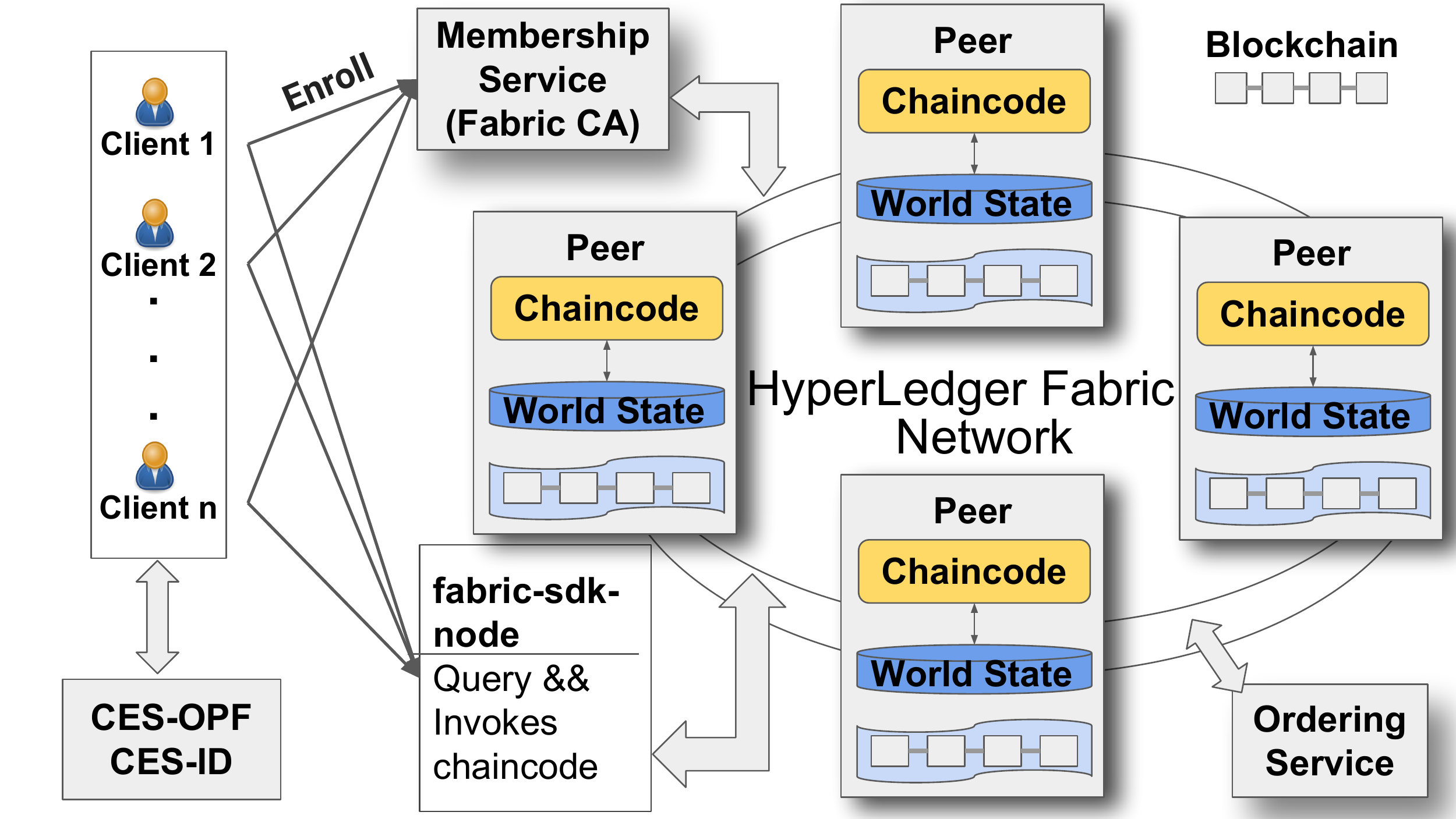}
		\vspace{-0.3cm}
	\caption{Architecture of combining Blockchain and smart contract with the optimization formulations presented in this paper.}
	\label{blockchain}
	\vspace{-0.64cm}
\end{figure}

\subsection{Blockchain Implementation using Hyperledger Fabric}
		
We implement Blockchain and smart contracts with the optimization as shown in Fig.~\ref{blockchain}. It consists of many nodes that communicate with each other, runs smart contracts called \texttt{chaincode}, holds state and ledger data. The clients shown are the end-users in the distribution network and can perform energy trading. Thousands of clients are allowed to connect to the Fabric network with minimal training. After enrolling in the network via \texttt{Fabric-CA}~\cite{worldstate}, a certificate needed for enrollment through a sdk, clients can communicate with a peer through \texttt{fabric-sdk-node}~\cite{worldstate}, update their preferences to Blockchain and store it in World State~\cite{worldstate} which is the database. Peers are used to commit transactions and maintain the world state and a copy of the ledger (consists of blocks). {The chaincode in Hyperledger Fabric is also deployed to peers and is executed as a user satisfies their commitments. Then, \textit{ordering service}, akin to mining in Bitcoin, generates new blocks in Fabric. Each peer will receive ordered state updates in the form of blocks from the ordering service. In this way, the order and number of blocks, a form of Blockchain, are maintained and synchronized for all peers.}

Algorithm~\ref{cecalgo} illustrates how the developed optimization routines are implemented with Blockchain and smart contracts. First, crowdsourcees communicate their preferences $\mc{X}$ with the utility or a system operator. Given these preferences, CES-OPF \eqref{equ:OPF} is solved for the market equilibrium. Smart contracts are then established for users $i \in \mc{G} \bigcup \mc{C}_{T_1}$ and are rewarded depending on their long-term contractual agreement with the utility. For Type 2 crowdsourcees operating in realtime markets, the CES-ID~\eqref{equ:incentives} is solved to obtain the monetary incentives communicated with the users. If client $i$ accepts the monetary incentive, then the transaction is finalized. However, if client $i$ refuses the incentive, the incentives are increased up to a certain budget. This can occur up to a certain threshold, where the operator can alternatively supplement the generation from traditional sources. The parameters $\eta_i$ and $\zeta_i$ for crowdsourcee $i$ are included to model their preferences and willingness to accept or reject the incentives.

\begin{algorithm}[h]
	\small	\caption{Blockchain-Assisted CES Operation}\label{cecalgo}
	\begin{algorithmic}[1]
\STATE Crowdsourcees input preferences and operational constraints $\mc{X}$
		\STATE  Solve CES-OPF~\eqref{equ:OPF}; obtain generator setpoints for $i \in \mc{G} \bigcup \mc{C}_{T1}$
		\STATE  Establish smart contracts for users $i \in \mc{G} \bigcup \mc{C}_{T1}$
		\STATE  Solve CES-ID~\eqref{equ:incentives} for generators $i \in \mc{C}_{T2}$; send incentives to Type 2 crowdsourcees with incentive $\m\lambda^a+\m \lambda$
		\IF{user $i \in \mc{C}_{T2}$  accepts incentives}
		\STATE  Finalize Blockchain transaction
		\ELSE
		\STATE Increase incentives for user $i$ up to a threshold/budget
		\STATE Update parameters ($\eta_i, \zeta_i$) to CES-ID~\eqref{equ:incentives} for user $i$
		\STATE If user $i$ rejects incentive again, supplement generation from traditional generators $i \in\mc{G}$
		\ENDIF
		\STATE Reconcile payments weekly or monthly
	\end{algorithmic}
\end{algorithm}

\section{Numerical Tests }\label{sec:tests}

We use the Southern California Edison 56-bus test feeder~\cite{peng2013optimal}. Reasonable load profile is generated to make sure the optimization problems have feasible sets for $T=24$ hrs. We place stationary batteries, solar panels, uncontrollable and shapable loads at each bus in the network. Batteries are set up with a power capacity of 80\% of the peak uncontrollable load at the bus, and a 4-hour energy storage capacity. We assume that the solar generation power profile is given and contributes to 150\% of the uncontrollable load at peak for each bus. Shapable loads have net energy demand that is up to 5 times the peak power consumption of the uncontrollable loads and can be charged for 4--10 hours. The self-scheduling time of shapable loads is between 9 am and midnight. A multi-period CES-OPF is implemented through ADMM as in~\cite{peng2014distributed}.

\begin{figure}[t]
	\centering \includegraphics[scale=0.2]{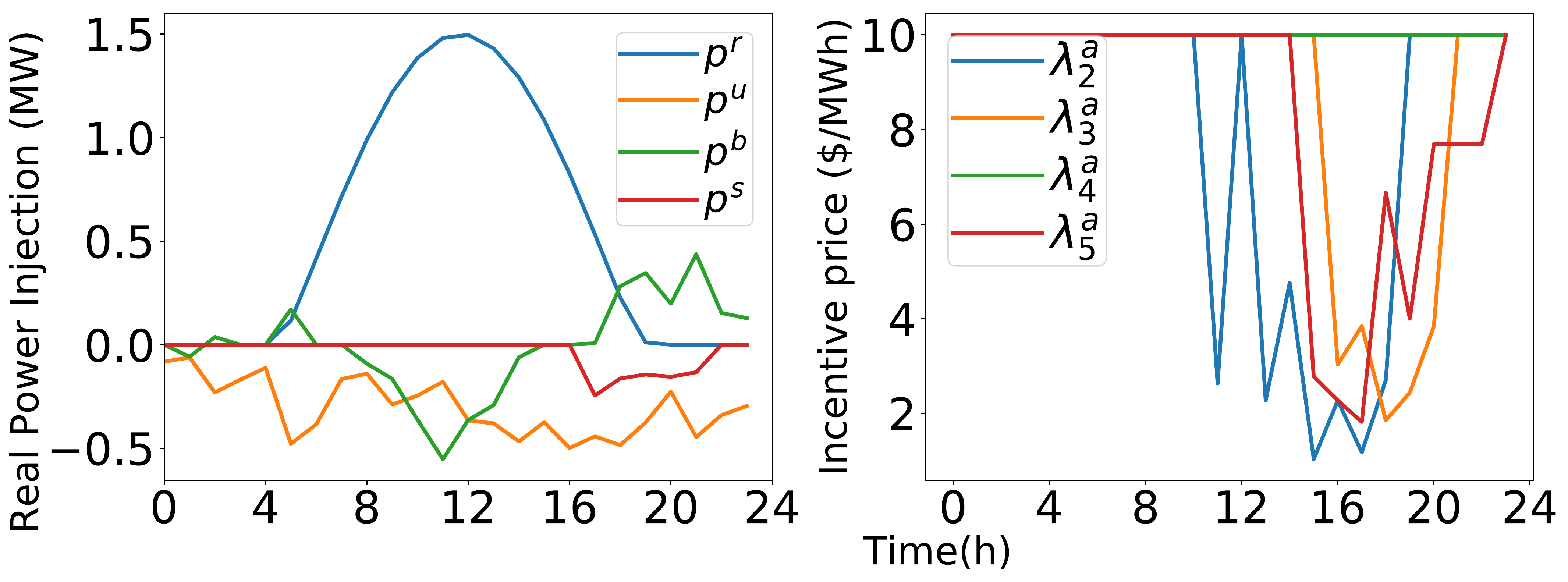}
	\caption{Load profile and generation from different sources for bus 5 (left), and crowdsourcing monetary incentives (right) for various buses.}
	\label{fig:OP1_PDF}
	\vspace{-0.35cm}
\end{figure}

\begin{figure}[t]
	\centering \includegraphics[scale=0.285]{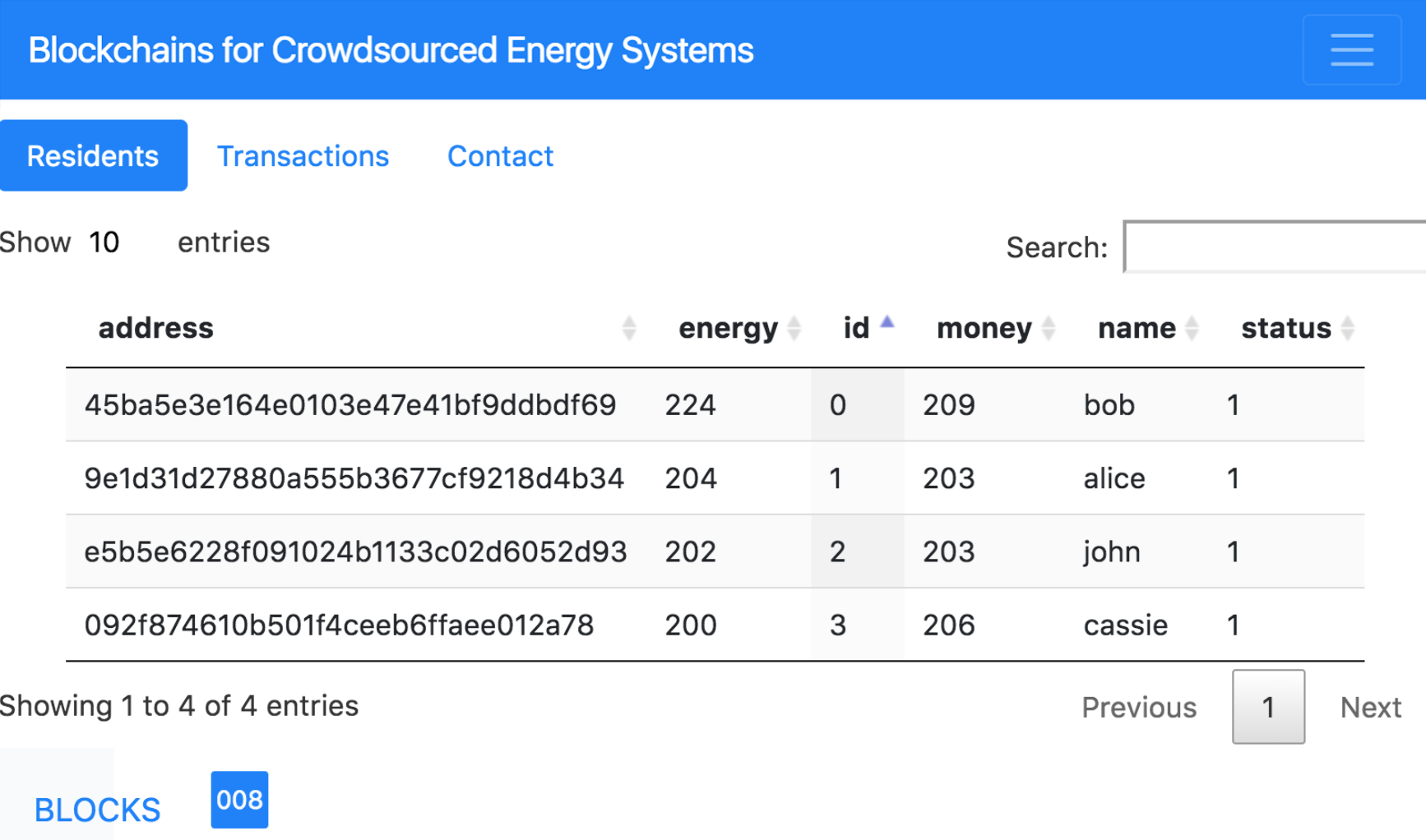}
	\vspace{-0.24cm}
	\caption{Web-based user interface for CESs with Hyperledger Fabric.}
	\label{fig:GUI.png}
	\vspace{-0.4cm}
\end{figure}

Fig.~\ref{fig:OP1_PDF} (left) shows the load profile and generation for bus 5 in the 56-bus power network after solving CES-OPF~\eqref{equ:OPF}. Due to space limitations, we only show 1 bus here. The figure also clearly shows that battery charges when the solar panel produces and injects energy into this bus. Fig.~\ref{fig:OP1_PDF} (right) presents the solutions to the designed incentives (i.e., the monetary rewards offered to Type 2 crowdsourcees) alongside simulating Algorithm~\ref{cecalgo}. The figure shows the time-varying nature of the incentives. Here, we assume that all crowdsourcees accept the designed incentives. Fig.~\ref{fig:GUI.png} shows a web-based user prototype that we implemented using Hyperledger Fabric.
In future work, we plan to study the convergence properties of the crowdsourcing incentives, and implement the framework on larger distribution networks. 

\vspace{-0.25cm}
\section*{Acknowledgments}
The authors acknowledge the anonymous reviewers and Dr. Nikolaos Gatsis for their helpful comments and suggestions.

\tiny
\bibliographystyle{IEEEtran}
\bibliography{IEEEabrv,bibfile}

\end{document}